\documentstyle [preprint,aps,epsf]{revtex}
\begin{document}
\title{Negative result measurements  in mesoscopic systems}
\author{S.A. Gurvitz\\
Department of Particle Physics, Weizmann Institute of
  Science\\ Rehovot 76100, Israel}
\vspace{18pt}
\maketitle

\begin{abstract}
We investigate measurement of electron transport in
quantum dot systems by using single-electron transistor
as a noninvasive detector. It is demonstrated that such a detector
can operate in the
``negative-result measurement'' regime. In this case   
the measured current is not distorted,
providing that it is a non-coherent
one. For a coherent transport, however, the
possibility of observing a particular state out of coherent
superposition leads to distortion of a measured
current even in the ``negative-result measurement'' regime.
The corresponding decoherence rate is obtained 
in the framework of quantum rate equations.
\end{abstract}

\hspace{1.5 cm}  
\vspace{18pt}

Rapid progress in nanoscale devices made it possible to
produce new type of detectors like the quantum point-contact
and the single electron transistor (SET), which have been already used
in different quantum measurements\cite{pep,mol,buks,dev}.
These devices have been considered also  
as possible detectors for a single two level system
(q-bit)\cite{g1,kor,shnir}.
It is of a great advantage that these detectors can be treated entirely
quantum mechanically, so that the related measurement process 
can be investigated in great details.
In particular, one can study quantum
mechanical mechanism of decoherence and its influence on a measured
system.

In this letter we consider a measurement of electron current
in quantum dots by using SET in close proximity of
a measured system, so it monitors the movement of single
electrons inside the system\cite{mol,dev}.
We demonstrate that varying parameters of SET one can put it 
in the ``negative result measurement'' regime\cite{ren}.
In this case the detector becomes a non-distractive if a measured
current is incoherent one. However, in the case of coherent measured current,
the negative result measurement distorts it via the 
decoherence. We evaluate the decoherence rate for this process  
and demonstrate that it is directly related to a possibility of
observation of a particular quantum state of the measured system
out of the linear superposition. Otherwise the negative
result measurement would not affect the measured current, even if
the latter is a coherent one. This phenomenon produces a peculiar
effect in the current which can be observed experimentally.    
 
We start with a description of measurement of
resonant tunneling currents in quantum dots by using SET. 
The system is shown schematically in Fig. 1\cite{mol}.
The SET, represented by the upper dot, is 
in close proximity to the lower dot (the measured system). 
Both dots are coupled to two separate reservoirs 
at zero temperature. The resonant levels $E_0$ and $E_1$ are taken between
the Fermi levels in the corresponding reservoirs,
$\tilde E_F^L>E_0>\tilde E_F^R$ and $E_F^L>E_1>E_F^R$.  
In the absence of electrostatic interaction between electrons  
the dc resonant currents in the detector and  
the measured system are respectively\cite{bg} 
\begin{equation}
I_D^{(0)}=e\frac{\gamma_L\gamma_R}{\gamma_L+\gamma_R},
~~~~~~~I_S^{(0)}=e\frac{\Gamma_L\Gamma_R}{\Gamma_L+\Gamma_R},
\label{a1}
\end{equation}
where $\gamma_{L,R}$ and $\Gamma_{L,R}$ are the tunneling 
partial widths of the levels 
$E_0$ and $E_1$ due to coupling with left and right 
reservoirs. The situation is different in 
the presence of electron-electron interaction between the dots, 
$H_{int}=Un_0n_1$, 
where $n_{0,1}$ are the occupancies of the upper and the lower dots
and $U$ is the Coulomb repulsion energy.  
If $E_{0}+U>\tilde E_F^L$, an electron from the left
reservoir cannot enter the upper dot when the lower dot is 
occupied [Fig.~1~(b)]. 
On the other hand, if an electron occupies the upper dot
[Fig.~1~(a$^{\prime}$,b$^{\prime}$)], the displacement of the level 
$E_1$  of the lower dot is less important, since it remains below 
the Fermi level, $E_{1}+U< E_F^L$. Thus, 
the upper dot can be considered as a detector  
registering charging of the lower dot via variation
of its current\cite{mol}. For instance, by measuring the variation of
the average detector current ($\Delta I_D$) 
due to the measurement, one can determine the average current
in the lower dot, $I_S$. If $\Delta I_D\gg I_S$, the detector represents
an amplifier, which can measure very small currents\cite{dev}.  
\vskip1cm
\begin{minipage}{13cm}
\begin{center}
\leavevmode
\epsfxsize=10cm
\epsffile{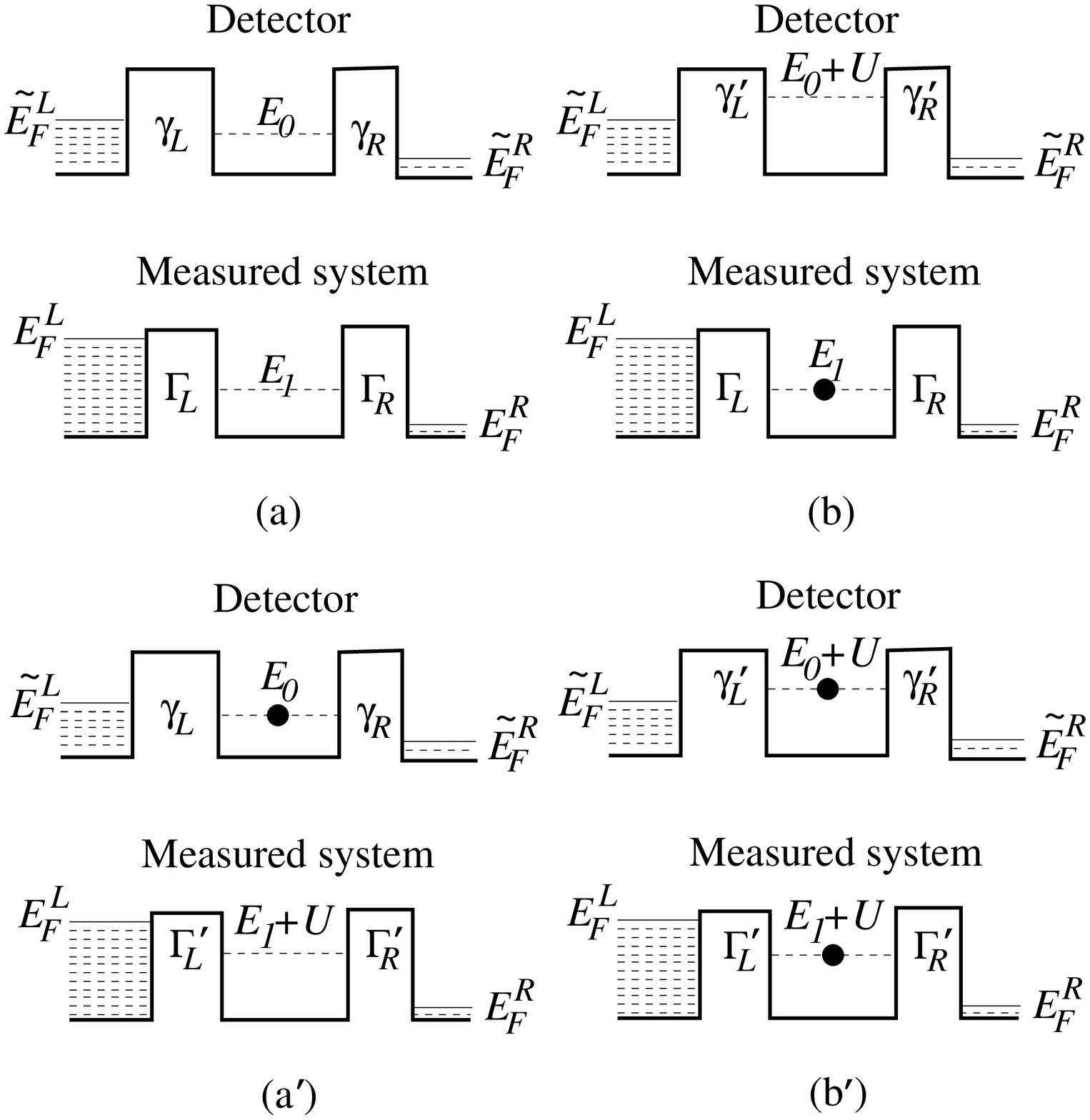}
\end{center}
{\begin{small}
{\bf Fig.~1:}
Measurement of resonant current in a single-dot structure by 
another, nearby dot. 
All possible electron states of the detector (the upper well)
and the measured system (the lower well) are shown. Also indicated are 
the tunneling rates ($\gamma$ and $\Gamma$) of the detector and the 
measured system respectively. 
\end{small}}
\end{minipage} \\ \\

In fact, the described detector affects the measured system. 
Indeed, if the detector is occupied 
[Fig.~1~(a$^{\prime}$,b$^{\prime}$)], an electron 
enters the lower dot with the energy $E_1+U$. As a result, the corresponding 
tunneling rates are modified ($\Gamma_{L,R}\to\Gamma'_{L,R}$), 
and therefore the measured current is distorted. One finds, however, 
that the states with empty detector, 
$|a\rangle$ and $|b\rangle$ [Fig.~1~(a,b)], 
do not distort the measured system. Nevertheless, the 
measurement process does take place:   
the detector current is 
interrupted whenever an electron occupies the measured system, but it 
flows freely when the measured system is empty. Such a measurement
is in fact the {\em negative} result
measurement\cite{ren}.  
Therefore, in order to put the above detector
in the negative result regime, we need to diminish 
the role of the states 
$|a'\rangle$ and $|b'\rangle$ in the measurement process.
It can be done by varying the penetrability of the detector barriers,
so that $\gamma_R\gg \gamma_L$. In this case  
the dwelling time of an electron inside the detector can be strongly
diminished. Indeed, the average charge inside 
a double-barrier structure\cite{bg} 
$e\gamma_L/(\gamma_L+\gamma_R )\to 0$ 
for $\gamma_L/\gamma_R\to 0$. Then    
an electron entering the detector leaves it immediately, 
remaining the measured system undistorted.   

Now we evaluate the measured current explicitly
in order to confirm that the above measurement is a non-distractive one.  
The currents through the detector and the measured system are 
determined by the density-matrix for the entire system $\rho (t)$,
which obeys the Schr\"odinger equation 
$i\dot\rho (t)=[{\cal H},\rho ]$ for ${\cal H}=H_D+H_S+H_{int}$, 
where $H_{D,S}$ are the tunneling Hamiltonians of the
detector and the measured system, respectively, and 
$H_{int}=Un_0n_1$. 
The current in the detector (or in the measured system) 
is the time derivative of the total average charge $Q(t)$ accumulated
in the corresponding right reservoir (collector): $I(t)=\dot Q(t)$, 
where $Q(t)=e$Tr$[\rho^R(t)]$ and $\rho^R(t)$ is the 
density-matrix of the collector. 
It was shown\cite{gp} that $I(t)$ is directly related to the 
density-matrix of the multi-dot system $\sigma_{ij}(t)$ with
$i,j=\{ a,a',b,b'\}$, 
obtained from the total density-matrix $\rho (t)$ by tracing out 
the reservoir states. 
One finds that the current in the detector or in the 
measured system is given by 
\begin{equation}  
I(t)=e\sum_j\sigma_{jj}(t)\Gamma^{(j)}_R, 
\label{a4}
\end{equation}
where the sum is taken over the states $|j\rangle$ in which the well 
adjacent to the corresponding collector is occupied, and 
$\Gamma_R^{(j)}$ is the partial width of the state $|j\rangle$
due to tunneling to the collector ($\gamma_R$ or $\Gamma_R$). 
In turn, $\sigma (t)$ obeys the following system of the rate 
equations \cite{gp}
\begin{mathletters}
\label{a2}
\begin{eqnarray}
&&\dot\sigma_{aa}  = -(\gamma_L+\Gamma_L)\sigma_{aa}
+\gamma_R\sigma_{a'a'}+\Gamma_R\sigma_{bb}
\label{a2a}\\
&&\dot\sigma_{bb}  = -\Gamma_R\sigma_{bb}+\Gamma_L\sigma_{aa}
+(\gamma'_L+\gamma'_R)\sigma_{b'b'}
\label{a2b}\\
&&\dot\sigma_{a'a'}  = -(\gamma_R+\Gamma'_L)\sigma_{a'a'}+\gamma_L\sigma_{aa}
+\Gamma'_R\sigma_{b'b'}
\label{a2c}\\
&&\dot\sigma_{b'b'}  = -(\gamma'_L+\gamma'_R+\Gamma'_R)\sigma_{b'b'}
+\Gamma'_L\sigma_{a'a'},
\label{a2d}
\end{eqnarray}
\end{mathletters}
where the states $|a\rangle$, $|b\rangle$, $|a'\rangle$, $|b'\rangle$
are the available states of the entire system, Fig. 1.
Note the off-diagonal density-matrix elements (coherencies) do not enter in
Eqs.~(\ref{a2}), and therefore these equations describe a non-coherent
transport. The reason is that there is no transitions between the isolate
states in this system, which characterize the coherent transport\cite{gp}. 

In the case of measurement, Fig.~1,
the currents in the detector and in the lower dot 
are $I^{(1)}_D(t)=e[\gamma_R\sigma_{a'a'}(t)+\gamma'_R\sigma_{b'b'}(t)]$
and $I^{(1)}_S(t)=e[\Gamma_R\sigma_{bb}(t)+\Gamma'_R\sigma_{b'b'}(t)]$,
respectively, Eq.~(\ref{a4}). 
The stationary (dc) current corresponds to $I=I(t\to\infty )$. 
Solving Eqs. (\ref{a2}) in the limit $\gamma_R,\gamma'_R\gg
\gamma_L,\gamma'_L$ we find 
\begin{equation}
\frac{\Delta I_D}{I^{(1)}_S}=\frac{\gamma_L}{\Gamma_R}\;,
~~~~~~~~~~~
I^{(1)}_S=e\frac{\Gamma_L\Gamma_R}{\Gamma_L+\Gamma_R}=I_S^{(0)}\;,
\label{a5}
\end{equation}
where $\Delta I_D=I^{(0)}_D-I^{(1)}_D$ is a variation of the detector
current with respect to the case of no measurement.
The first equation shows that the SET amplifies quantum
signals if the ratio $\gamma_L/\Gamma_R\gg 1$. Thus,  
one can measure small current $I_S$ by measuring  
variation of the detector current $\Delta I_D$\cite{pep}.    
On the other hand the measured current $I_S$ is not distorted 
by the detector, as follows from the second equation. 

Consider now a measurement of resonant transport in a coupled-dot
structure\cite{vdr}, Fig. 2. In this case the electron transport
is a coherent one, since an electron
inside the double dot appears in the linear
superposition of two states ($E_1$ and $E_2$).
The SET detector, represented by
the upper dot, is taken in close proximity to the second dot
of the double-dot system, thus measuring charging of that dot.  
For simplicity, we assume strong Coulomb repulsion between two
electrons inside the coupled-dot, so only one electron can occupy  
the measured system\cite{naz}. Fig.~2 shows all possible electron
configurations of the double-dot when the detector is empty.
Similar to the previous case (Fig.~1) each of the states,
$|a\rangle$, $|b\rangle$, $|c\rangle$ has its 
counterpart $|a'\rangle$, $|b'\rangle$, $|c'\rangle$
corresponding to the occupied detector. 
$U_{1,2}$ is the Coulomb repulsion
energy between the detector and the measured system for the 
electron occupying the first or the second dot.
We consider $E_0+U_2> \tilde E_F^L$, but $E_0+U_1< \tilde E_F^L$.
Therefore 
the detector is blocked only when the second dot is occupied.
\vskip1cm
\begin{minipage}{13cm}
\begin{center}
\leavevmode
\epsfxsize=6cm
\epsffile{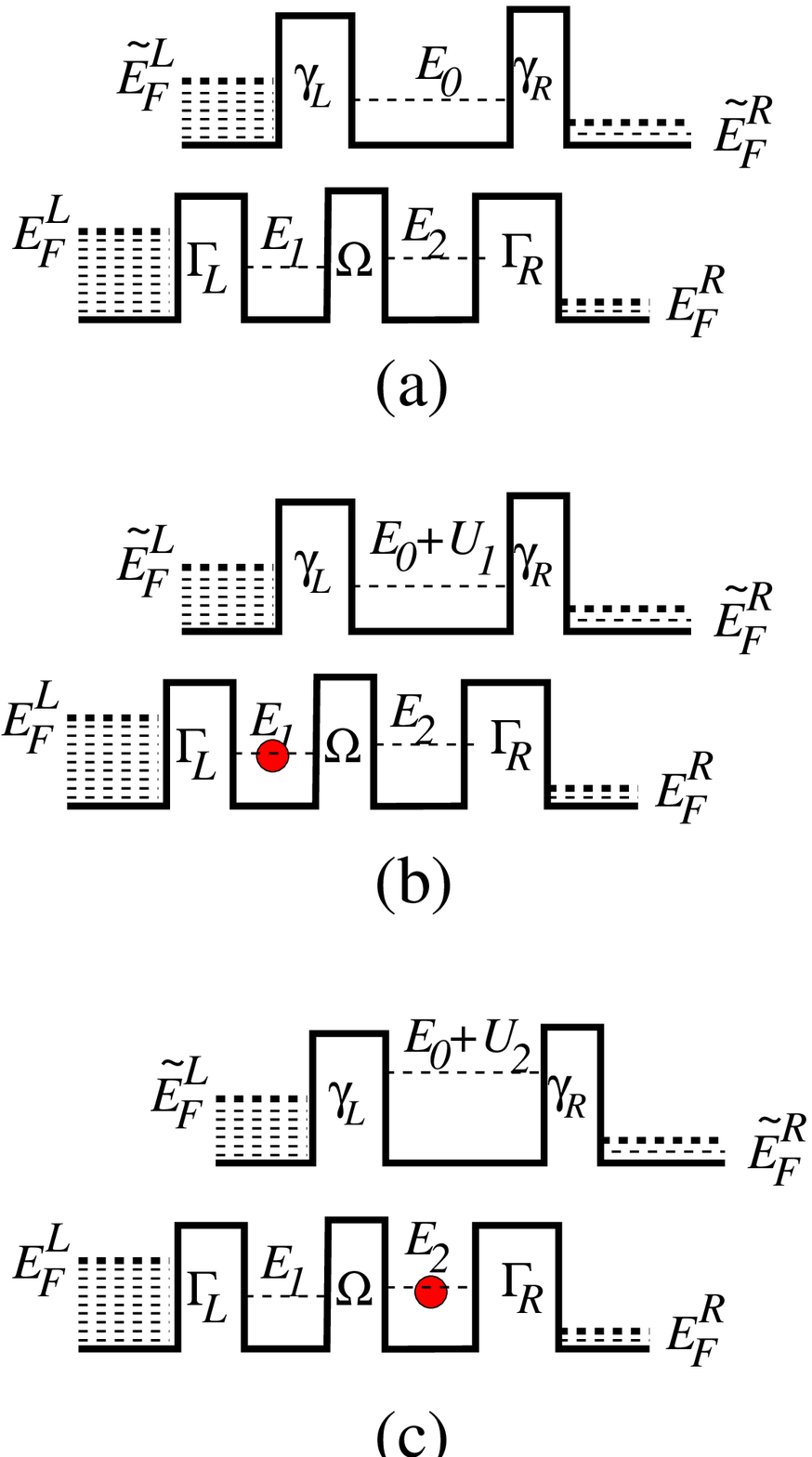}
\end{center}
{\begin{small}
{\bf Fig.~2:}
Measurement of resonant current in a double-dot structure.
Only the states with empty detector (the upper dot) are shown.
\end{small}}
\end{minipage} \\ \\

Consider first the case of no measurement (no interaction with the
upper dot). The available states of the double-dot system, 
$|a\rangle$, $|b\rangle$ and $|c\rangle$, are those  
as shown in Fig.~2. 
The resonant current through this system 
is described by the Bloch-type rate equations, derived from
the microscopic Schr\"odinger equation\cite{gp}
\begin{mathletters}
\label{a3}
\begin{eqnarray}
&&\dot\sigma_{aa}  = -\Gamma_L\sigma_{aa}+\Gamma_R\sigma_{cc}
\label{a3a}\\
&&\dot\sigma_{bb}  = \Gamma_L\sigma_{aa}+i\Omega (\sigma_{bc}-\sigma_{cb})
\label{a3b}\\
&&\dot\sigma_{cc}  = -\Gamma_R\sigma_{cc}-i\Omega (\sigma_{bc}-\sigma_{cb})
\label{a3c}\\
&&\dot\sigma_{bc}  = i\epsilon\sigma_{bc}+i\Omega (\sigma_{bb}-\sigma_{cc})
-\frac{1}{2}\Gamma_R\sigma_{bc},
\label{a3d}
\end{eqnarray}
\end{mathletters}
where $\epsilon =E_2-E_1$ and $\sigma_{cb}=\sigma_{bc}^*$.  
The diagonal density-matrix elements $\sigma_{ii}$ are 
the probabilities of finding the system 
in one of the states, $|a\rangle$, $|b\rangle$ 
and $|c\rangle$. In the distinction with  
the resonant tunneling through a single dot, 
the diagonal density-matrix elements are coupled with the 
non-diagonal elements $\sigma_{bc}$, $\sigma_{cb}$ (``coherences'')
by a hopping amplitude between two isolated 
states, $E_1$ and $E_2$\cite{gp}. 
The total resonant dc flowing through 
this system is $I_S^{(0)}=e\Gamma_R\sigma_{cc}(t\to\infty )$,
Eq.~(\ref{a4}).
Solving Eqs.~(\ref{a3}) one obtains\cite{naz}  
\begin{equation}
I^{(0)}_S=e\frac{\Gamma_R\Omega^2}{\epsilon^2+\Gamma^2_R/4
+\Omega^2(2+\Gamma_R/\Gamma_L)}
\label{a7}
\end{equation}
Note that the dissipation of the ``coherences'', $\sigma_{bc}$,  
is generated by the last term in
Eq.~(\ref{a3d}), proportional to the half of decay rates of 
the states $|b\rangle$ and  $|c\rangle$ due to their coupling with 
the reservoirs. (In our case the state $|b\rangle$ cannot decay, 
but only the state $|c\rangle$, so the corresponding dephasing rate    
is proportional to $\Gamma_R$). Since 
the resonant current proceeds via hopping between two dots, 
generated by $\sigma_{bc}$, it decreases with $\Gamma_R$, Eq.~(\ref{a7}).

Now we ``switch on''the detector. 
The available states of the entire system are 
$|a\rangle$, $|b\rangle$, $|c\rangle$, Fig. 2, and 
$|a'\rangle$, $|b'\rangle$, $|c'\rangle$, corresponding to 
empty and occupied detector, respectively. For simplicity we assume
that all transition tunneling amplitudes are weakly dependent on
energy, so $\Gamma ,\gamma ,\Omega =\Gamma' ,\gamma' ,\Omega' $,  
The rate equations describing the transport in the entire system are\cite{gp}  
\begin{mathletters}
\label{a6}
\begin{eqnarray}
&&\dot\sigma_{aa} =-(\Gamma_L+\gamma_L)\sigma_{aa}+\gamma_R\sigma_{a'a'}
+\Gamma_R\sigma_{cc}
\label{a6a}\\
&&\dot\sigma_{a'a'} =-(\Gamma_L+\gamma_R)\sigma_{a'a'}
+\gamma_L\sigma_{aa}+\Gamma_R\sigma_{c'c'}
\label{a6b}\\
&&\dot\sigma_{bb} =\Gamma_L\sigma_{aa}+i\Omega (\sigma_{bc}-\sigma_{cb})
-\gamma_L\sigma_{bb}+\gamma_R\sigma_{b'b'}
\label{a6c}\\
&&\dot\sigma_{b'b'} = \Gamma_L\sigma_{a'a'}+i\Omega
(\sigma_{b'c'}-\sigma_{c'b'})
+\gamma_L\sigma_{bb}-\gamma_R\sigma_{b'b'}
\label{a6d}\\
&&\dot\sigma_{cc} = -\Gamma_R\sigma_{cc}-i\Omega
(\sigma_{bc}-\sigma_{cb})+(\gamma_L+\gamma_R)\sigma_{c'c'}
\label{a6e}\\
&&\dot\sigma_{c'c'} = -\Gamma_R\sigma_{c'c'}
-i\Omega (\sigma_{b'c'}-\sigma_{c'b'})-(\gamma_L+\gamma_R)\sigma_{c'c'}
\label{a6f}\\
&&\dot\sigma_{bc} = i\epsilon\sigma_{bc}+i\Omega (\sigma_{bb}-\sigma_{cc})
-\frac{1}{2}(\Gamma_R+\gamma_L)\sigma_{bc}
+\gamma_R\sigma_{b'c'}\\
\label{a6g}
&&\dot\sigma_{b'c'} = i(\epsilon-U_1+U_2)\sigma_{b'c'}
+i\Omega (\sigma_{b'b'}-\sigma_{c'c'})
-\frac{1}{2}(\gamma_L+2\gamma_R+\Gamma_R)\sigma_{b'c'}\, .
\label{a6h}
\end{eqnarray}
\end{mathletters}
Solving these equations one finds the 
average current in the detector and the coupled-dot system, given by Eq.~(\ref{a4}):
$I^{(1)}_D=e\gamma_R(\sigma_{a'a'}+\sigma_{b'b'}+\sigma_{c'c'})$, 
and $I^{(1)}_S=e\Gamma_R(\sigma_{cc}+\sigma_{c'c'})$. 

Let us take again the limit of the negative result measurement,  
$\gamma_R\gg\gamma_L$, in which the detector is  
not expected to affect the measured system. If so,  
the density-matrix of the entire system, 
traced over the detector states would coincide 
with the density-matrix for the double-dot system without detector, 
Eqs.~(\ref{a3}). However, this is not the case. Indeed, by 
introducing the reduced density matrix of the measured system,
$\bar\sigma_{ij}=\sigma_{ij}+\sigma_{i'j'}$, 
one finds from Eqs.~(\ref{a6}) that in the above limit of the negative
result measurement $\bar\sigma_{ij}$ obeys the 
following equations
\begin{mathletters}
\label{a8}
\begin{eqnarray}
&&\dot{\bar\sigma}_{aa}  = -\Gamma_L\bar\sigma_{aa}+\Gamma_R\bar\sigma_{cc}
\label{a8a}\\
&&\dot{\bar\sigma}_{bb}  = \Gamma_L\bar\sigma_{aa}+i\Omega (\bar\sigma_{bc}-\bar\sigma_{cb})
\label{a8b}\\
&&\dot{\bar\sigma}_{cc}  = -\Gamma_R\bar\sigma_{cc}-i\Omega (\bar\sigma_{bc}-\bar\sigma_{cb})
\label{a8c}\\
&&\dot{\bar\sigma}_{bc}  = i\epsilon\bar\sigma_{bc}+i\Omega (\bar\sigma_{bb}-\bar\sigma_{cc})
-\frac{1}{2}(\Gamma_R+\gamma_L)\bar\sigma_{bc},
\label{a8d}
\end{eqnarray}
\end{mathletters}
where the resonant current flowing through this system is
respectively 
$I_S^{(1)}=e\Gamma_R\bar\sigma_{cc}(t\to\infty )$, Eq.~(\ref{a4}).
Note that Eqs.~(\ref{a8}) obtained for the SET detector in the
negative result regime, coincide with the rate equations for the
point-contact detector\cite{g1}, although the both detectors operate in a 
different way. 

Let us compare Eqs.~(\ref{a8}) with Eqs.~(\ref{a3}). 
One finds that equations for the diagonal matrix 
elements are the same. Yet it is not so for 
the off-diagonal matrix elements. The difference is in the 
additional dephasing rate, $\gamma_L/2$, generated by the detector. 
It is easy to trace its origin. In accordance with the Bloch 
equations the dissipation of the nondiagonal density-matrix elements 
$\bar\sigma_{bc}$ is the half of all possible decay rates of each 
of the states ($|b\rangle$ and $|c\rangle$). In the presence 
of the detector, the state $|b\rangle$, Fig. 2, has  
an additional decay channel, corresponding to the possibility for an 
electron to enter the detector. Thus, despite of   
the dwelling time of an electron in the detector tends to zero and
therefore the related detector state does not distort the measured 
system, the {\em possibility} for an electron to enter the detector  
influences the measured current 
very drastically. Indeed, solving Eqs.~(\ref{a6}), (\ref{a8}) one obtains 
\begin{equation}
\frac{\Delta I_D}{I^{(1)}_S}=\frac{\gamma_L}{\Gamma_R}\;,
~~~~~~~~~~~
I^{(1)}_S=e\frac{\Gamma_R\Omega^2}{\epsilon^2/\eta +\eta\Gamma^2_R/4
+\Omega^2(2+\Gamma_R/\Gamma_L)}\not =I^{(0)}_S\, ,
\label{a9}
\end{equation}
where $\eta =1+(\gamma_L/\Gamma_R)$. If we compare Eqs.~(\ref{a9})
with Eqs.~(\ref{a5}) we find that SET can measure the resonant current
in a coupled-dot structure precisely in the same way
as in the previous case of a single dot. 
However, the measured system is distorted now. For instance,  
if $\gamma_L\Gamma_R\gg \Omega^2$ and $\epsilon =0$, 
the measured current $I_S^{(1)}\simeq I^{(0)}_S/\eta\ll I^{(0)}_S$. 

The additional decoherence rate $\gamma_L$ appears in Eq.~(\ref{a8d}) 
only when the detector can {\em distinguish} 
a particular dot occupied by an electron. 
Yet, such an ``observation'' effect disappears if   
$\tilde E_F^L<E_0+U_1$. In this case an electron cannot 
enter the detector no matter which of the dots of the measured 
system is occupied. Then the additional decay channel for
the state $|c\rangle$ is blocked and 
Eq.~(\ref{a8d}) coincides with 
Eq.~(\ref{a3d}), i.e. the measured average current remains undistorted, 
$I_S^{(1)}=I_S^{(0)}$ although the detector still interacts with the measured
system ($\Delta I_D\not =0$). Such a peculiar dependence of the 
average current $I_S$ on $\tilde E_F^L$ is shown in Fig. 3. This ``measurement''
effect can be observed experimentally by varying the detector voltage,
or by moving the resonance level $E_0$.    
\vskip1cm
\begin{minipage}{13cm}
\begin{center}
\leavevmode
\epsfxsize=6cm
\epsffile{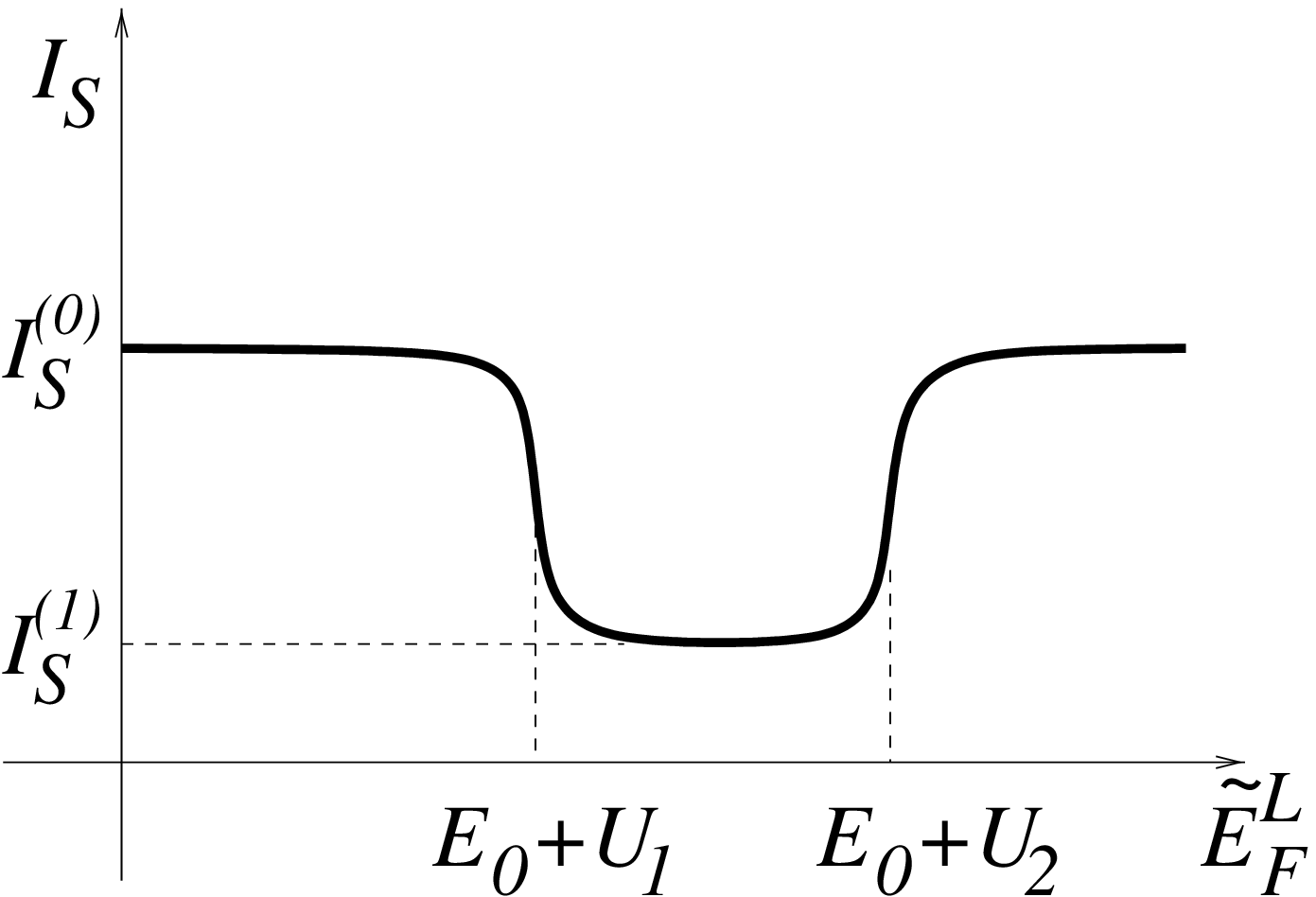}
\end{center}
{\begin{small}
{\bf Fig.~3:}
The average current in the double-dot structure with aligned level ($E_1=E_2$)
as a function of the Fermi energy of the 
left reservoir adjacent to the detector.
\end{small}}
\end{minipage} \\ \\

In conclusion, by using the quantum rate equations
method we demonstrated that SET detector can work in 
the negative result measurement regime. In this case the SET detector 
does not distort an observed system, if
its motion is determined by classical rate equations,
i.e. involving only the diagonal density-matrix elements. 
Such a situation is realised in the resonant transport through
a single level. 
However, in the case of coherent transport, as in the resonant
tunneling through coupled dots, the
negative result measurement always distorts a measured system,
providing that the detector can measure the charging of a particular dot.
This measurement effect is accounted by the quantum rate equation,
which allows us to evaluate the corresponding decoherence rate. 
However, if the detector cannot distinguish the charging of a particular dot
it becomes a non-distractive again.  

Parts of this work were done while the author stayed at 
University of Trento, Trento, Italy and TRIUMF, Vancouver, Canada. I thank 
these institutions for their hospitality. I am also  
grateful to A. Korotkov for important comments, 
related to this work.

\end{document}